\documentclass{iopart}

\usepackage{epsfig}
\usepackage{here}
\usepackage{cite}

\begin{document}
  \title[]{A method to disentangle single- and
       multi-meson production
       in missing mass spectra
       from quasi-free ${\bf{pn\to pn X}}$ reactions
  }

\author{P.~Moskal$^{1,2}$\footnote{e-mail address: p.moskal@fz-juelich.de},
        H.-H.~Adam$^3$, A.~Budzanowski$^4$, R.~Czy{\.z}ykiewicz$^1$, D.~Grzonka$^2$,
        M.~Janusz$^1$, L.~Jarczyk$^1$, T.~Johansson$^5$, B.~Kamys$^1$,
        P.~Klaja$^1$, A.~Khoukaz$^3$, K.~Kilian$^2$, J.~Majewski$^1$,
        W.~Oelert$^2$, C.~Piskor-Ignatowicz$^1$, J.~Przerwa$^1$,
        J.~Ritman$^2$, T.~Ro{\.z}ek$^{2,6}$, T.~Sefzick$^2$,
        M.~Siemaszko$^6$, J.~Smyrski$^1$, A.~T{\"a}schner$^3$,
        J.~Wessels$^3$, P.~Winter$^2$, M.~Wolke$^2$, P.~W\"ustner$^2$,
        Z.~Zhang$^2$, W.~Zipper$^6$
       }

\address{ $^1$ Nuclear Physics Department, Jagellonian University, 30-059 Cracow, Poland}
\address{ $^2$ IKP \& ZEL Forschungszentrum J\"ulich, 52425 J\"ulich, Germany}
\address{ $^3$ IKP, Westf\"alische Wilhelms-Universit\"at, 48149 M\"unster, Germany}
\address{ $^4$ Institute of Nuclear Physics, 31-342 Cracow, Poland}
\address{ $^5$ Department of Radiation Science, Uppsala University, 
          75121 Uppsala, Sweden}
\address{ $^6$ Institute of Physics, University of Silesia, 40-007 Katowice, Poland}

\begin{abstract}
  The separation of contributions from multi- and  single-meson
  production in the missing mass spectrum of the quasi-free $pn\to
  pn X$ reaction constitutes a~challenging task when the reaction is
  studied close to threshold. This is especially true if the
  resolution of the mass determination is comparable with the excess
  energy and if the investigated signal appears close to the
  kinematical limit. In this article we outline a method which
  permits the extraction of the signal originating from the creation
  of a single meson without the necessity of conducting
  model-dependent simulations. For the $pd\to pnXp_{spectator}$
  reactions, the method allows one to combine events corresponding
  to multi-meson production at various excess energies with respect
  to the $pn\to pn~meson$ process, and hence leads to an increase of
  the statistics needed for the determination of the shape of the
  multi-meson background.

  As an example of the application of the method, we demonstrate
  that the evaluation of the data from the $pd\to pnXp_{sp}$ process
  according to the described technique enables one to extract a
  signal of the $pn\to pn\eta$ reaction whose shape is consistent
  with expectations, supporting the correctness and usefulness of
  the method introduced.
\end{abstract}


\submitto{\JPG}

\section{Introduction}
\label{Introduction} In recent years, after many very successful
studies of the close-to-threshold production of mesons in
proton-proton collisions~\cite{wilkin,hanhart,review,machner,wilkin0},
several experimental groups have extended the range of their
investigations to proton-neutron
scattering~\cite{calen1,calen2,calen3,bilgerspectator,wasa,tof,anke1,anke2,hadron,joannaleap,wasaatcosy}.
The results obtained can then be compared with theoretical models
of the production of mesons in different isospin channels and this
reduces significantly the ambiguities of such models. However, the
experiments and the evaluation of the data from the proton-neutron
interaction are much more difficult than measurements of
proton-proton reactions. This is mainly due to having to use
nuclear targets as a source of neutrons. In order to define the
full kinematics of the $pn\to pnX$ events one needs to measure at
least one more particle than for the analogous meson production in
proton-proton scattering. This is because the determination of the
four-momentum vector of the colliding quasi-free neutron requires
the measurement of the non-interacting residue of the target
nucleus (spectator) or the decay products of the created meson. In
comparison to the $pp\to pp$~\emph{meson} reactions, the
efficiency for registering the quasi free $pn\to pn$~\emph{meson}
process is reduced by a large factor and, additionally, the
resolution in the four-momentum of the investigated meson also
worsens significantly. Therefore, in the evaluation of the data
one faces problems of low statistics, which are especially
important when the analysis demands subtraction of histograms and
when the expected signal is close to the kinematical limit. Such
is the situation when the quasi-free $pn\to pn$~\emph{meson}
reaction is investigated by registering all the outgoing nucleons
from the $pd\to p_{sp} pnX$ reaction, using the missing mass
technique to identify the desired meson events. The subscript $sp$
denotes the spectator proton which, it is assumed, does not take
part in the reaction and $X$ stands for the single meson or
multi-meson system.

An identification of the $pn\to pn$~\emph{meson} reaction on an
event-by-event basis is impossible if only the nucleons are
measured. One can, however, extract the number of registered
$pn\to pn$~\emph{meson} events from the missing mass distribution
provided that the contribution of the continuous spectrum
originating from multi-pion production can be subtracted from the
signal.

In this article, we describe a method for the subtraction of a
physical multi-pion background from the missing mass spectrum for
quasi-free $pn\to pn X$ reactions, where the decay products of the
produced meson are not detected. Such a technique is clearly
required in, e.g., the derivation of the correct number of $pn\to
pn\eta$ and $pn\to pn\eta^{\prime}$ events from the missing mass
spectra obtained by the COSY-11
collaboration~\cite{hadron,joannaleap,michalproc}.

In the next section we introduce the method in the idealized case
where the number of events is infinite. This simplification will
allow us to focus first on the problem of the proximity of the
signal to the kinematical limit. After explaining the main ideas
of the background subtraction, we will generalize the discussion
to the situation encountered in real experimental conditions.
Thus, in section~\ref{estimation} we will demonstrate how in the
case of the low statistics one can use all events taken at
negative excess energies and combine them to construct the shape
of the background.
Section~\ref{justification} will be devoted  to the justification
of the assumptions made and finally in section~\ref{example} we
will show an example of the application of the technique to data
taken at the COSY-11 facility.

Though the method was developed for the analysis of the COSY-11
$pd\to pn\eta p_{sp}$ results, the conclusions are valid also for
the production of the $\eta^{\prime}$ and other narrow mesons.
Moreover, since the technique is independent of the detection
system, it may be also applied when evaluating data obtained at
other facilities.

\section{Derivation of the background assuming infinite statistics}
\label{derivation}%
The COSY-11 facility used to investigate the quasi-free $pn\to pn
X$ reactions is equipped with detectors for the registration of
fast charged particles~\cite{brauksiepe,NIMjurek},
neutrons~\cite{joannamgr,joannameson}, and slow spectator
protons~\cite{bilgerspectator}. Experiments are conducted using
the COSY proton beam~\cite{cosy1,cosy2} incident on the internal
deuteron cluster target~\cite{dombrowski,khoukaz}.

The total C.M.\ energy $\left(\sqrt{s}\,\right)$ available for the
proton-neutron scattering is fixed by the four-vectors of the
incident and spectator protons. Due to the internal motion of the
nucleons in the deuteron, this varies from event to event and in
the $\eta$ region a range of about 60~MeV is scanned for a fixed
beam momentum~\cite{hadron,c11,stepaniak}. 
The four-momentum vectors of the beam and spectator protons are known
for each registered event.
Therefore, the collected
data can be grouped according to the excess energy with respect to
the $pn\to pn\eta$ process. The production of the $\eta$ meson can
only occur if the excess energy $Q$,
\begin{equation}
  \label{excess}
  Q = \sqrt{s} - m_{proton} - m_{neutron} - m_{\eta}
\end{equation}
is positive. For negative $Q$ only pions can be created.

  \begin{figure}[h]
       \parbox{1.0\textwidth}{\centerline{
           \epsfig{file=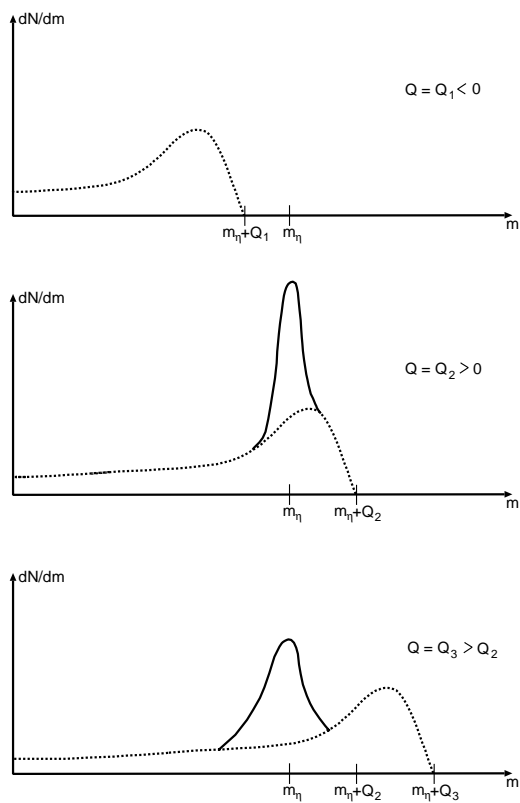,width=0.52\textwidth,angle=0}}}\\
   \parbox{1.0\textwidth}{\caption{ \small
             Schematic depiction of the missing mass spectra
             for the $pn\to pn X$ reaction below ($Q=Q_1$)
             and above ($Q=Q_2$ and $Q=Q_3$)
             the threshold for the $\eta$ meson production.
             The broadening of a signal from the $\eta$ meson with the increasing
             of $Q$ is kinematical effect discussed in detail in reference~\cite{jureketa}.
           \label{missscheme}
         }}
  \end{figure}
Fig.~\ref{missscheme}  shows in a schematic manner the mass
distributions expected at three different values of $Q$ for the
unobserved system $X$ produced via the $pn\to pn X$ reaction. The
upper panel presents an outlined missing mass spectrum when $Q<0$.
The shape of this continuous distribution arises from a
convolution of the real invariant mass distribution of the
multi-pion system, the efficiency of the detection setup for the
simultaneous registration of nucleons from the $pn\to
pn$~\emph{pions} reaction, and also from the experimental
resolution of the missing mass determination. For $Q>0$ a signal
from the $\eta$ meson is expected on top of the multi-pion mass
distribution at a position corresponding to its mass, as shown in
the middle panel of Fig.~\ref{missscheme}. For data at higher $Q$
the peak from the $\eta$ meson will remain at the same position on
the mass axis, though the distribution of multi-pion invariant
mass will spread towards larger masses (lowest panel of
Fig.~\ref{missscheme}), extending up to the kinematical limit.
However, in practice, as will be argued later, the change of the
multi-pion background can be approximated as a shift of the entire
spectrum in the direction of higher masses.

In the case of the $pn\to pn\eta$ reaction measured at the COSY-11
facility, the experimental mass resolution amounted to about
5~MeV(FWHM)~\cite{rafalmgr}. Therefore, for measurements at only
few MeV above the threshold the shape of the pion background must
be well known in order to extract the $\eta$ signal. The lack of a
clear indication of the upper end of the signal 
resulting from the $\eta$ meson production
(middle panel of Fig.~\ref{missscheme})
constitutes the main problem in disentangling the contributions
from the $\eta$ meson and pions.

For simplicity of argumentation, let us first assume in
subtracting the multi-pion background that the number of
registered events is very large, such that one can neglect any
statistical fluctuations, and that the shape of the reconstructed
invariant mass of the multi-pion background is independent of the
excess energy $Q$. The form of the background could then be
determined from a missing mass spectrum $dN^{\pi}/{dm}$ from an
infinitesimal range at any negative values $Q$. Having postulated
that the shape of the background $dN^{\pi}(m,Q)/dm$ does not
depend on $Q$, it is convenient to express it in terms of a
normalised function $B(m_{\eta}+Q-m)$ of the difference between
the kinematical limit ($m_{\eta} + Q$) and the given mass $m$.

\begin{figure}[H]
       \parbox{1.0\textwidth}{\centerline{
                \epsfig{file=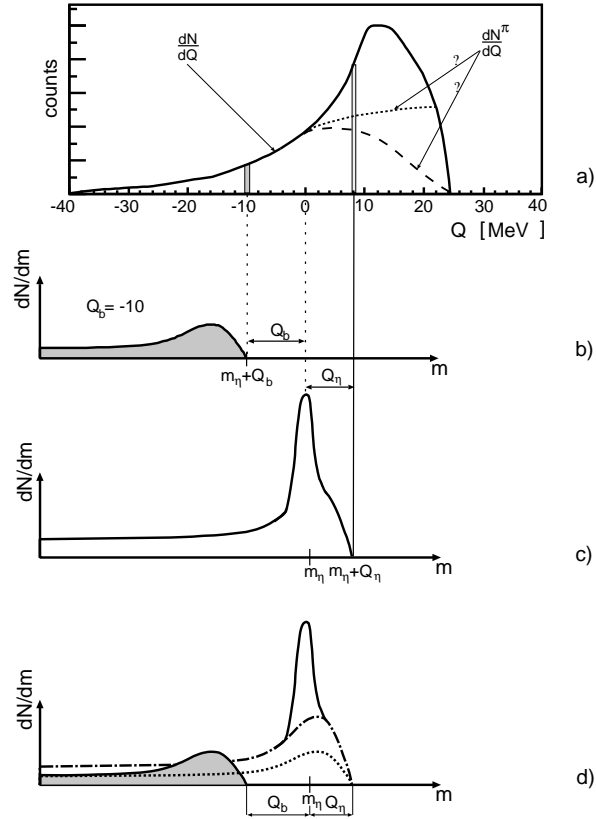,width=0.6\textwidth,angle=0}}}\\
   \parbox{1.0\textwidth}{\caption{ \small
           a) Distribution of the excess energy with respect to the $pn\to pn\eta$
              reaction plotted schematically for a beam momentum of $P_b~=~2.075$~GeV/c.
              The shape results from i) the genuine excitation functions for $\eta$ meson
              and multi-pion production in the proton-neutron collisions, ii) the distribution
              of the  Fermi momentum
              of the nucleons in the deuterium target,
              and iii) the acceptance and efficiency of the detection system.
              For a detailed discussion of this issue see e.g.\ ref.~\cite{hadron,c11}.
              The contribution to the spectrum from pion production~$dN^{\pi}/{dQ}$
              cannot be derived unambiguously from this spectrum. Therefore,
              entirely arbitrary,  two
              different  possibilities
              are pointed out as dotted and dashed curves.\protect\\
           b), c) Form of the missing mass spectra as derived for the negative ($Q_b$) and positive
              ($Q_\eta$) values of $Q$.   To each value of $Q$ a
              continuous spectrum of masses can be assigned,
              originating e.g.\ from the  $pn\to pn\pi\pi$ reaction.
              However, there is always a unique relation between the
              value of $Q$ and the maximum mass which can be created ($m_{max}~=~m_{\eta}~+~Q$).
              Therefore, for better visualisation,
              the axis of $Q$ for the ${dN}/{dQ}$ distribution
              and the mass axis of the missing mass spectra ${dN}/{dm}$ were arranged such that
              a maximum mass achievable for a given value of $Q$ lies exactly below the corresponding
              $Q$ value in the ${dN}/{dQ}$ plot.\protect\\
           d) The shape of the background in the mass spectrum for $Q>0$
              can be constructed from the shape of the multi-pion mass distribution (shaded histogram)
              after its shift
              to the kinematical limit (dotted line)  and subsequent
              normalisation at low masses (dash-dotted line).
           \label{Qmiss}
         }}
  \end{figure}

Being not limited by statistics, one could divide the positive $Q$
range into very narrow subranges within which the resultant
missing mass spectrum would be the simple sum of a signal from the
$\eta$ meson and the form $B$ multiplied by an appropriate factor.
The discussed situation is depicted schematically in
Fig.~\ref{Qmiss}, where the lower panel shows the method of how to
construct the background.

If the two assumptions mentioned above were valid then, in order
to derive a signal of the $\eta$ meson from a missing mass
spectrum for positive Q, it would be sufficient to subtract from
this spectrum a missing mass distribution  determined for negative
$Q$ after the shift of the latter to the kinematical limit~(dotted
line)  and normalisation  at the very low mass values where no
events from the $\eta$ meson production are expected~(dash-dotted
line). In such a case the contribution of the $pn\to pn\eta$
reaction could be extracted without the necessity of any additional 
assumption regarding the unknown distribution of the background
expressed as a function of the excess energy ${dN^{\pi}}/{dQ}$. In
this case the analysis would not demand an exact knowledge of the
dynamics of the background reactions.
\newpage

\section{Background estimation under the real experimental conditions}
\label{estimation}

Due to the finite statistics, one needs to study the missing mass
spectrum in finite intervals of $Q$. This does not influence the
shape of the missing mass distribution arising from the $pn\to pn
\eta$ reaction, though it alters the form of the multi-pion
background. The latter ${dN^{\pi}}/{dm}(m)$, when derived from the
finite $\Delta Q$ range, becomes the convolution of the $B$
function and the number of background events expressed as a
function of the excess energy ${dN^{\pi}}/{dQ}$. For the range
spanned between $Q_{min}$ and $Q_{max}$ the missing mass
distribution of the background can be expressed as:
\begin{equation}
 \frac{dN^{\pi}}{dm}(m) = \int_{Q_{min}}^{Q_{max}}
 \frac{dN^{\pi}}{dQ}(Q)~B(m_{\eta} + Q - m)~dQ.
  \label{equation2}
\end{equation}
In the $dN/dQ$ spectrum we cannot extract the function
${dN^{\pi}}/{dQ}$ in a model-independent way for positive values
of $Q$ (compare dashed and dotted lines in Fig.~\ref{Qmiss}a). We
must therefore take as narrow ranges of $\Delta{Q}$ as reasonable,
taking into account the experimental resolution in $Q$, and to
assume that within such a range the ${dN^{\pi}}/{dQ}$ function is
constant. It is natural to take the width of the $Q$ subranges
equal to the FWHM of the resolution of the $Q$ determination. Such
a width choice partially validates the constant ${dN^{\pi}}{dQ}$
assumption, due to the unavoidable experimental smearing within
the bin.

\begin{figure}[h]
\centerline{\epsfig{file=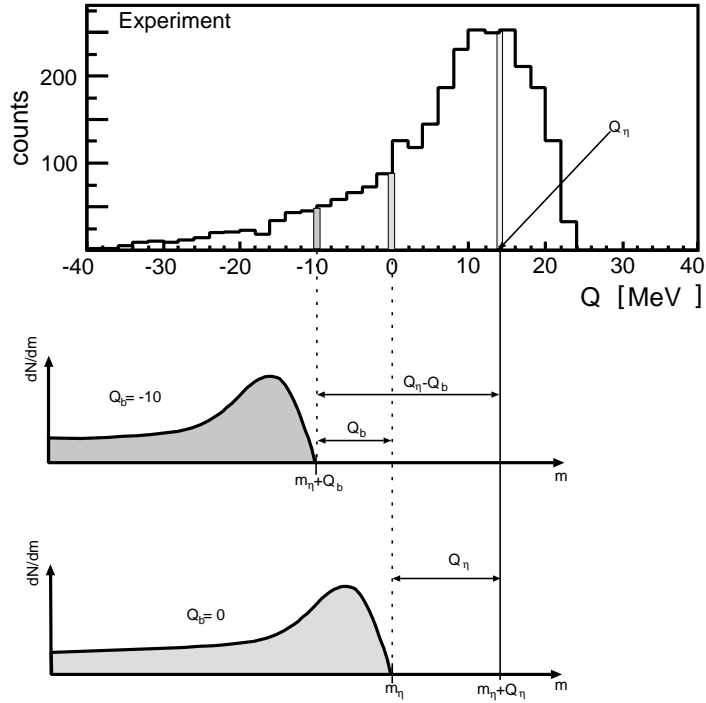,width=0.72\textwidth,angle=0}}
\vspace{-0.3cm} \caption{\label{mmpisum}
          Upper panel: Experimental distribution of the excess energy~$Q$ as
          measured with the COSY-11 setup, using a deuterium target and proton beam
          with a momentum of 2.075~GeV/c~\cite{hadron}.\protect\\
          Lower panels: Scheme of the missing mass distribution for two arbitrarily chosen
          values of $Q\le 0$.
         }
\end{figure}

By reviewing the experimental distribution ${dN}/{dQ}$ shown in
the upper panel of Fig.~\ref{mmpisum}, one recognizes that the
obtained statistics are indeed insufficient to derive the
${dN^{\pi}}/{dm}$ spectrum from the bin of $Q$ with a width equal
to the experimental resolution (i.e.\ FWHM~=~5~MeV). The
statistics could, however, be improved significantly if all events
registered with negative $Q$ could be taken into account. This can
be realized by adding to the missing mass calculated for a given
event an amount ($Q_{\eta}~-~Q_{b}$), which will shift the
background events measured at $Q_{b}$ such as if they were
determined at the excess energy of $Q_{\eta}$. The resultant
modified missing mass distribution obtained from the entire data
sample of negative $Q$ values can be identified with the function
$B(m_{\eta}+Q_{\eta}-m)$ needed for the derivation of the
background distribution  within a finite excess energy bin
centered around the $Q_{\eta}$ value. In such a way, when
investigating the missing mass spectrum for the $Q_{\eta}$ bin,
one would obtain directly the background spectrum which, after the
normalisation, could be subtracted from the distribution
containing the signal from the $\eta$ meson. When the number of
events is small, this approach enables one to reduce significantly
the statistical fluctuations in the background spectrum. However,
as was already stressed, the entire procedure could only be
justified if the shape of the background did not alter within the
discussed range of the excess energy $Q$. Arguments in favour of
this assumption will be presented in the subsequent section.

\section{Why does the experimental multi-pion background not change significantly with
         variation of the excess energy?}
\label{justification}

We now present arguments supporting the supposition that the
measured shape of the background does not alter significantly with
changes in excess energy, at least over the energy range studied.
It is important to stress that this assumption worked correctly
when evaluating the data of the reactions $pp\to
pp\eta^{\prime}$~\cite{etap} and $pp\to pp\eta$~\cite{jureketa}
and that the derived cross sections are consistent with those
obtained at the SATURNE facility~\cite{hibou}.

As already mentioned, the shape of the background spectrum reflects
i) the real distribution of the invariant mass of the multi-pion systems,
ii) the acceptance of the detection setup,
and iii) the experimental resolution.
The acceptance of the detection system, its efficiency and
resolution do not change significantly over the considered
variation of excess energy.
On the level of a few per cent, this is true even in the case of
the detection setups characterized by a very limited geometrical
acceptance, such as e.g.\ the COSY-11 facility~\cite{f0}.
The shape of the genuine background distribution is the only
significant factor that is relevant to our discussion.
In practice, from the high statistics measurements of the $pp\to
pp\eta$ reaction, we know that the shape of the background could
be sufficiently well described assuming that it was due to
two-pion creation only~\cite{hab,prc,eduard}. Adding into the
simulations three- and four-pion events was superfluous, and did
not improve the agreement with the experimental distributions
significantly.
  \begin{figure}[H]
       \vspace{-0.2cm}
       \parbox{1.0\textwidth}{\centerline{
                \epsfig{file=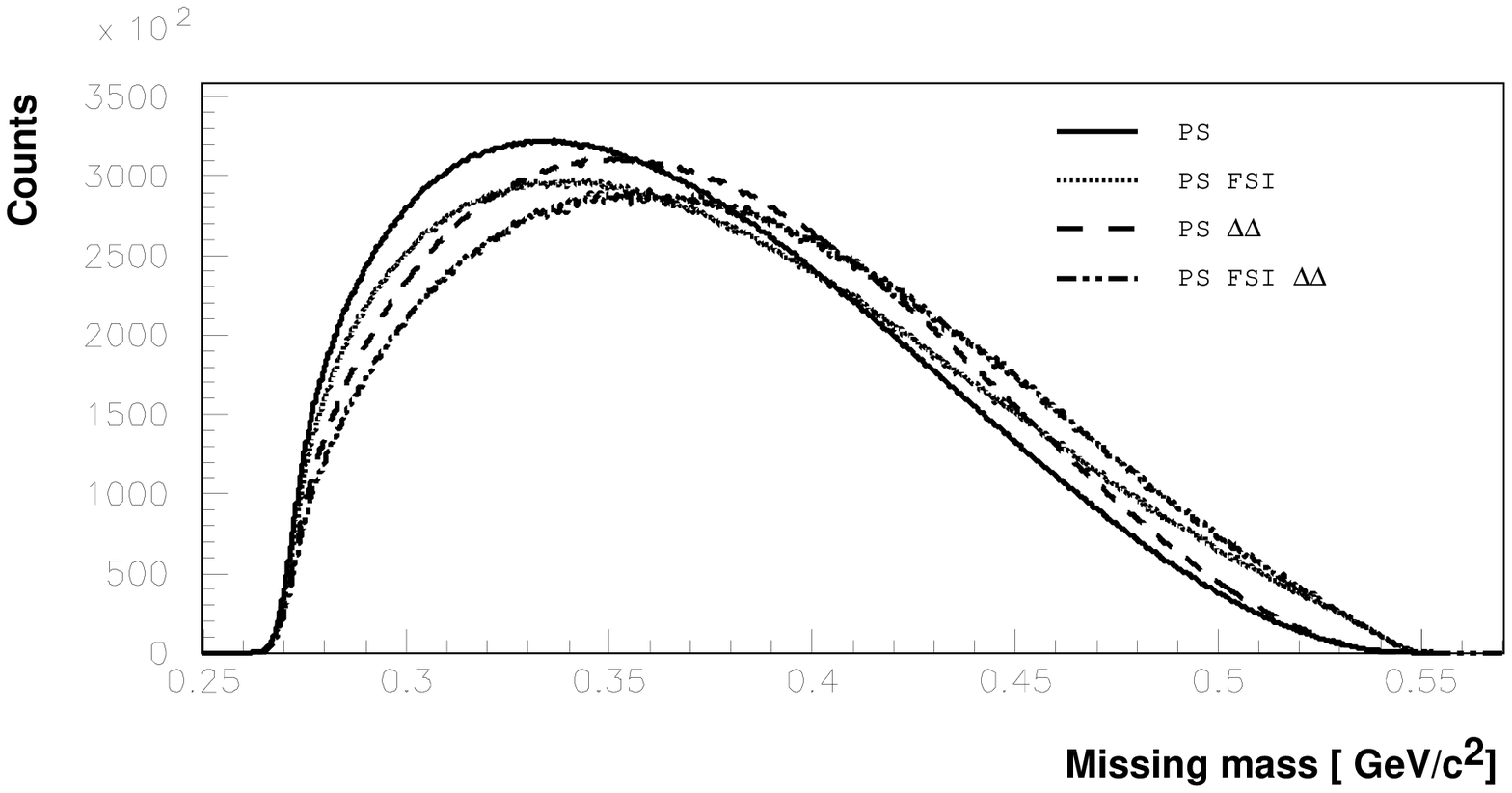,width=0.7\textwidth,angle=0}}}\\
       \vspace{-0.4cm}
       \parbox{1.0\textwidth}{\centerline{
                \epsfig{file=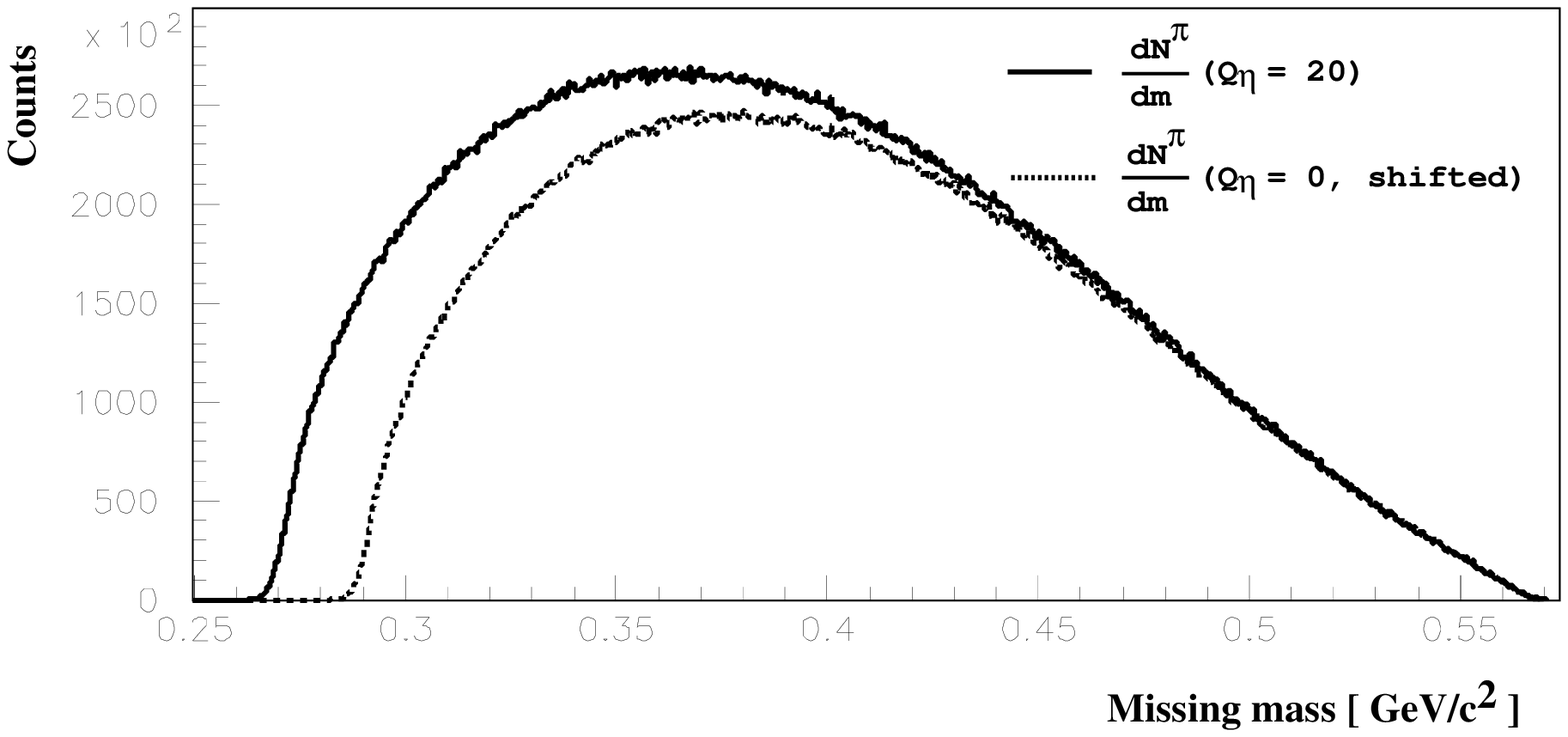,width=0.7\textwidth,angle=0}}}\\
       \vspace{0.4cm}
       \parbox{1.0\textwidth}{\centerline{
                \epsfig{file=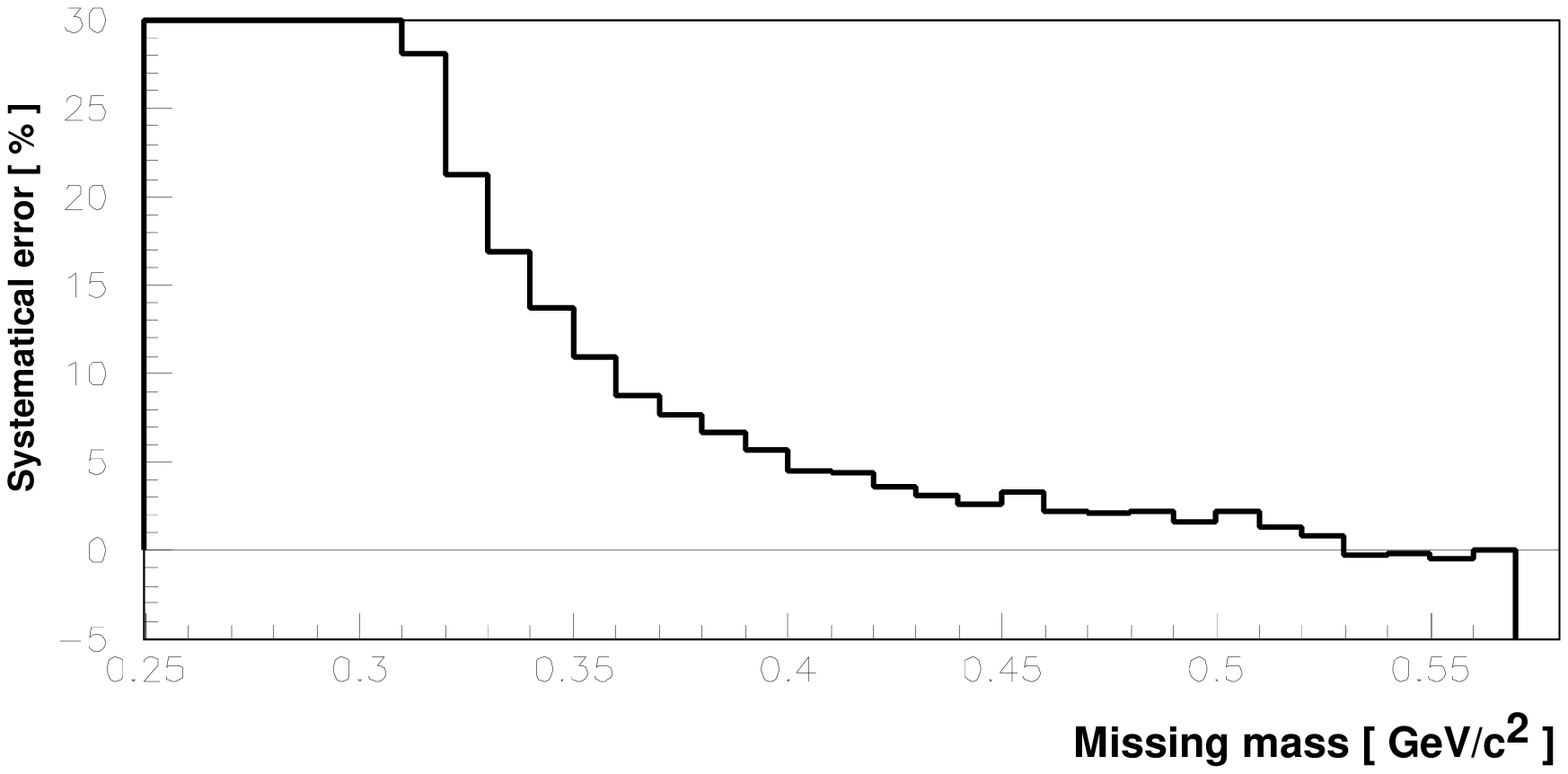,width=0.67\textwidth,angle=0}}}\\
       \vspace{-0.4cm}
   \parbox{1.0\textwidth}{\caption{ \small
           (Upper panel) Missing mass with respect to the 
           $pn$ system calculated for the $pn\to pn \pi\pi$ process 
           at the threshold for the $pn\to pn \eta$ reaction.
           Solid line shows uniform phase-space distribution,
           dotted line presents its modification  
           by the final state interaction (FSI) between proton and neutron,
           and dashed line  represents spectrum obtained assuming that the pions are produced via the
           $pn\to \Delta\Delta \to p \pi n \pi$ reaction chain.
           Dashed-dotted line shows the effects of both the proton-neutron FSI
           and the resonance production together. \\
           (Middle panel) 
           Comparison between the distributions calculated regarding resonance production
           and FSI for $Q_{\eta}$~=~20 MeV (solid line) and $Q_{\eta}$~=~0~MeV (dotted line).
           The latter is shifted by 20~MeV.  \\
           (Lower panel) Difference between the spectra of the middle panel normalised to the solid line:\\
            $(dN^{\pi}(Q_{\eta}=20)/dm - dN^{\pi}(Q_{\eta}=0)/dm )~/~(dN^{\pi}(Q_{\eta}=20)/dm)$.
           \label{monte}
         }}
  \end{figure}
Fig.~\ref{monte} demonstrates the
missing mass distribution of $\pi\pi$ system
produced via the  $pn\to pn X$ reaction
calculated taking into account experimental resolution of the detection system and
various creation scenarios. 

  \begin{figure}[H]
       \vspace{-0.2cm}
       \parbox{1.0\textwidth}{\centerline{
                \epsfig{file=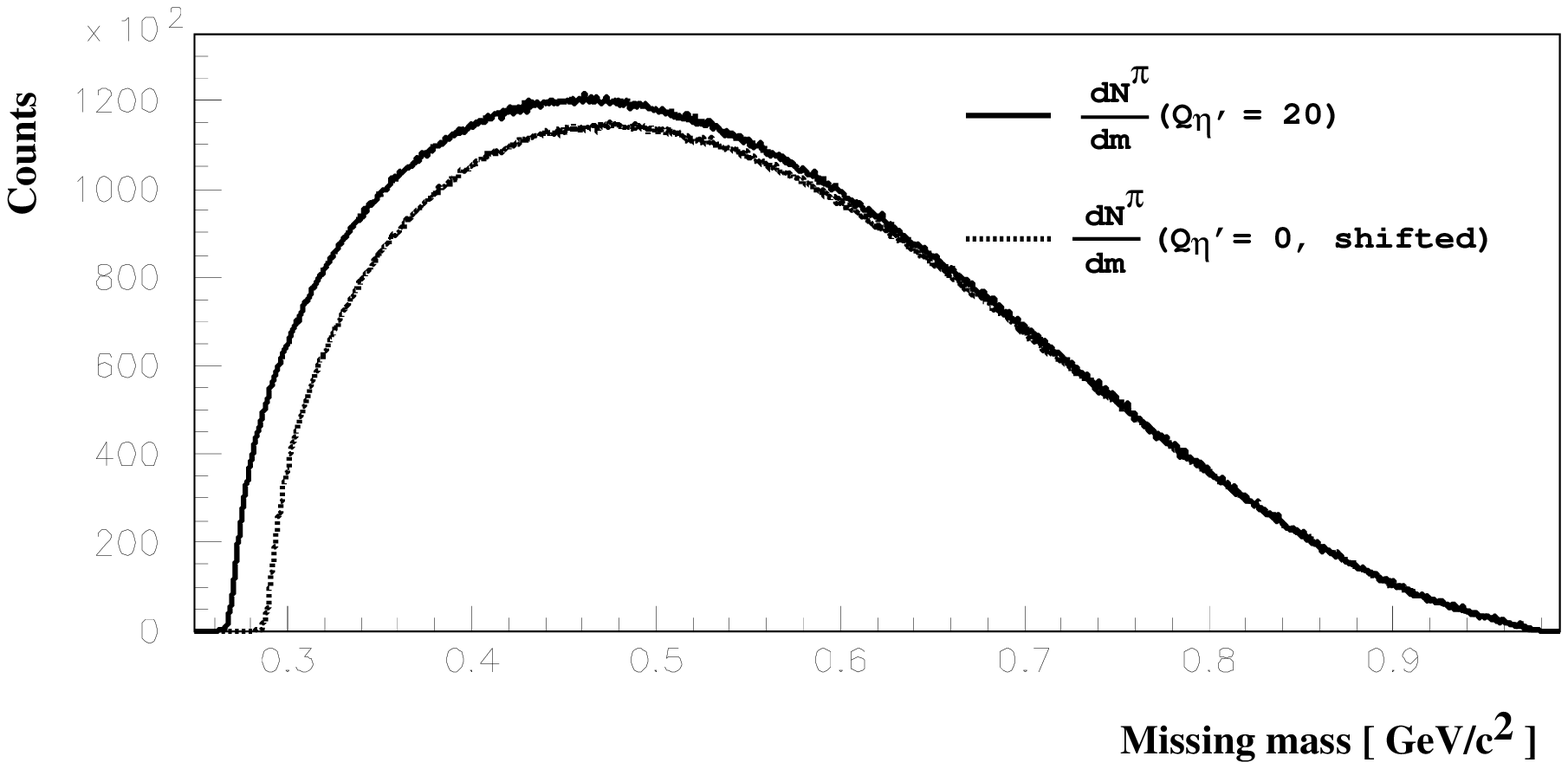,width=0.7\textwidth,angle=0}}}\\
       \vspace{-0.4cm}
       \parbox{1.0\textwidth}{\centerline{
                \epsfig{file=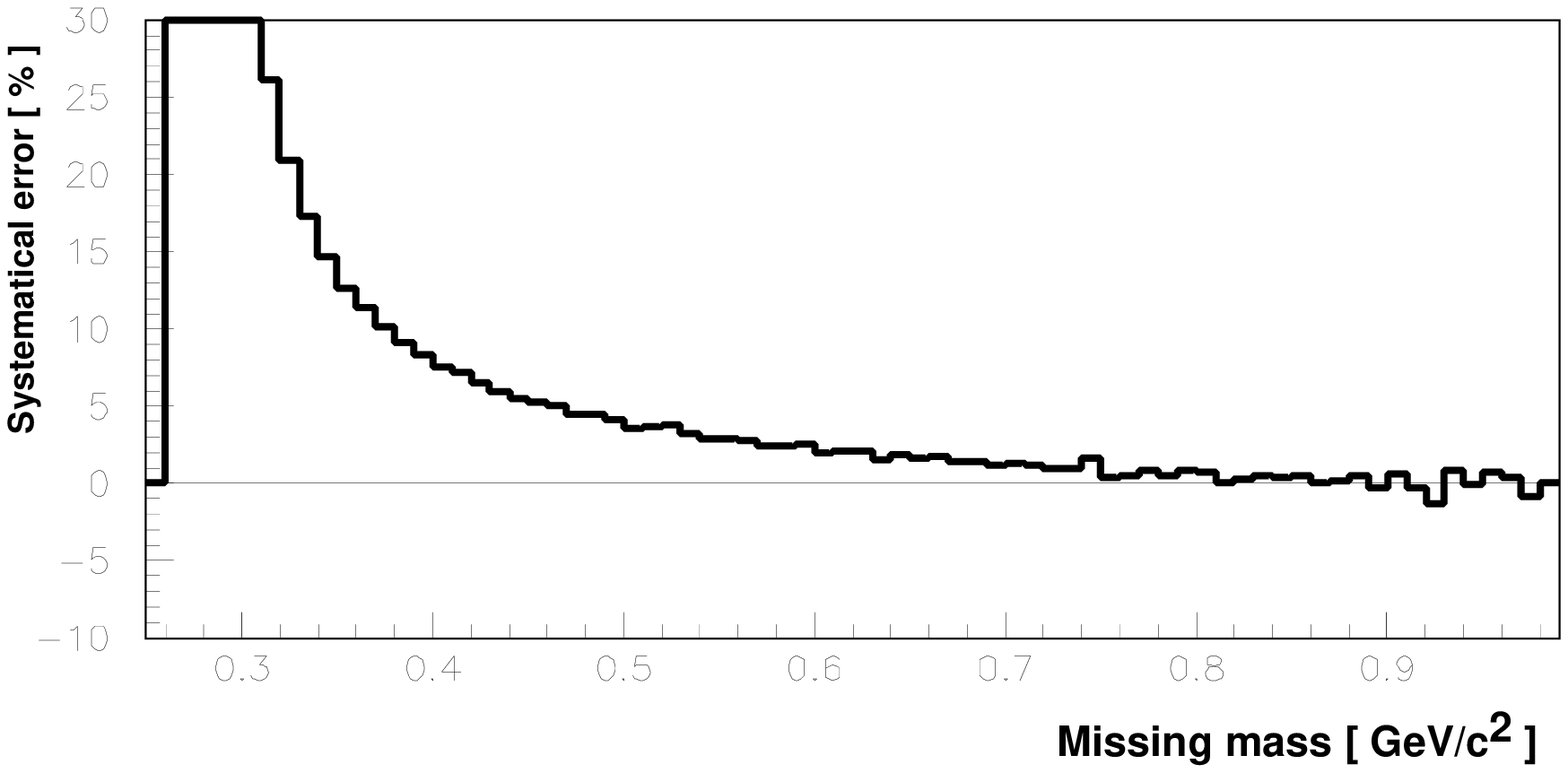,width=0.7\textwidth,angle=0}}}\\
   \parbox{1.0\textwidth}{\caption{ \small
           (Upper panel) Missing mass with respect to the 
           $pn$ system calculated for the $pn\to pn \pi\pi$ process 
           at the threshold (dotted line) and 20 MeV above the 
           threshold (solid line) of $pn\to pn \eta^{\prime}$  reaction.
           The dotted histogram was shifted by 20 MeV towards the larger masses.
           (Lower panel) Difference between the spectra of the upper  panel normalised to the solid line:
     $(dN^{\pi}(Q_{\eta^{\prime}}=20)/dm - dN^{\pi}(Q_{\eta^{\prime}}=0)/dm )~/~(dN^{\pi}(Q_{\eta^{\prime}}=20)/dm)$.
           \label{monte5}
         }}
  \end{figure}

When $Q$ increases the distribution of the multi-pion invariant
mass broadens towards larger masses. In practice, if the change in
the value of $Q$ is small compared to the excess energy with
respect to threshold of the $pn\to pn~pions$ reaction, the shape
of the mass distribution at its edge remains approximately
unaltered. Of interest for us is the range of about 40~MeV/c$^2$
from the higher mass limit. The events represented by the solid
and dotted lines in the middle panel of Fig.~\ref{monte} were simulated at excess
energies of Q~=~20 and Q~=~0~MeV with respect to the $pn\eta$
system. From the comparison of the form of both spectra one
realises that at the edges there is no noticeable difference
between the shape of the curves. Thus, this shape, convoluted with
an acceptance function independent of the $Q$ value, will lead to
the form of the background which will be a function only of the
distance from the kinematical limit, over at least a range of
about 40~MeV. 

The quantitative appraisals of the systematical error caused by the subtraction of the shifted
and normalized spectra is depicted in the lower panel of figure~\ref{monte}.
This picture demonstrates a fractional systematical
error  due to the discussed method.
It was calculated as a difference between the mass distributions
$dN/dm^{\pi}$ at Q~=~20~MeV and Q~=~0~MeV normalized to the $dN/dm^{\pi}$(Q~=~20~MeV).
From the figure one can infer that the systematical error in the missing mass range pertinent
to our discussion
(close to the kinematical limit) is in the order of a few per cent of the background value.
In the middle and lower panels of Fig.~\ref{monte} we show the comparison for the production model depicted by
the dashed-dotted line in the upper panel,
however, we have checked that for other discussed possibilities the determined systematical error is similar.

It is worth noting that in the case of the
$\eta^{\prime}$ meson, due to its twice larger mass, the relative
changes between dotted and solid lines are even smaller (see upper panel of Fig.~\ref{monte5}), and
hence the advocated assumption is even better justified.
    The lower panel of Fig.~\ref{monte5} indicates that the fractional error due to the changes in the shape
     is in the order of one per cent only within the range of 200~MeV in the vicinity of the kinematical limit.

\section{Example of application}
\label{example}

\begin{figure}[H]
       \parbox{0.14\textwidth}{\mbox{}}
       \parbox{0.43\textwidth}{\centerline{
                \epsfig{file=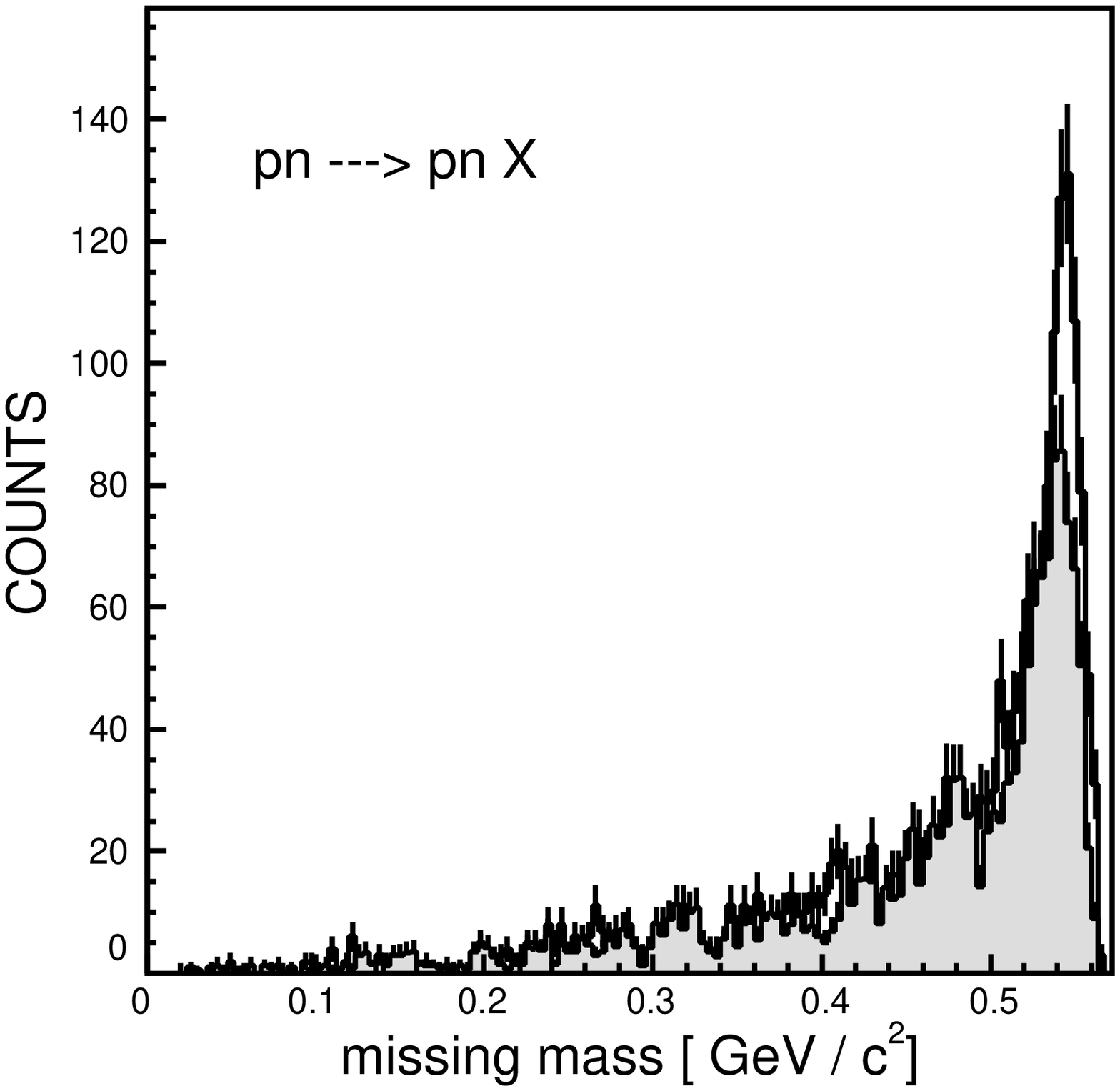,width=0.42\textwidth,angle=0}}}
       \parbox{0.43\textwidth}{\centerline{
                \epsfig{file=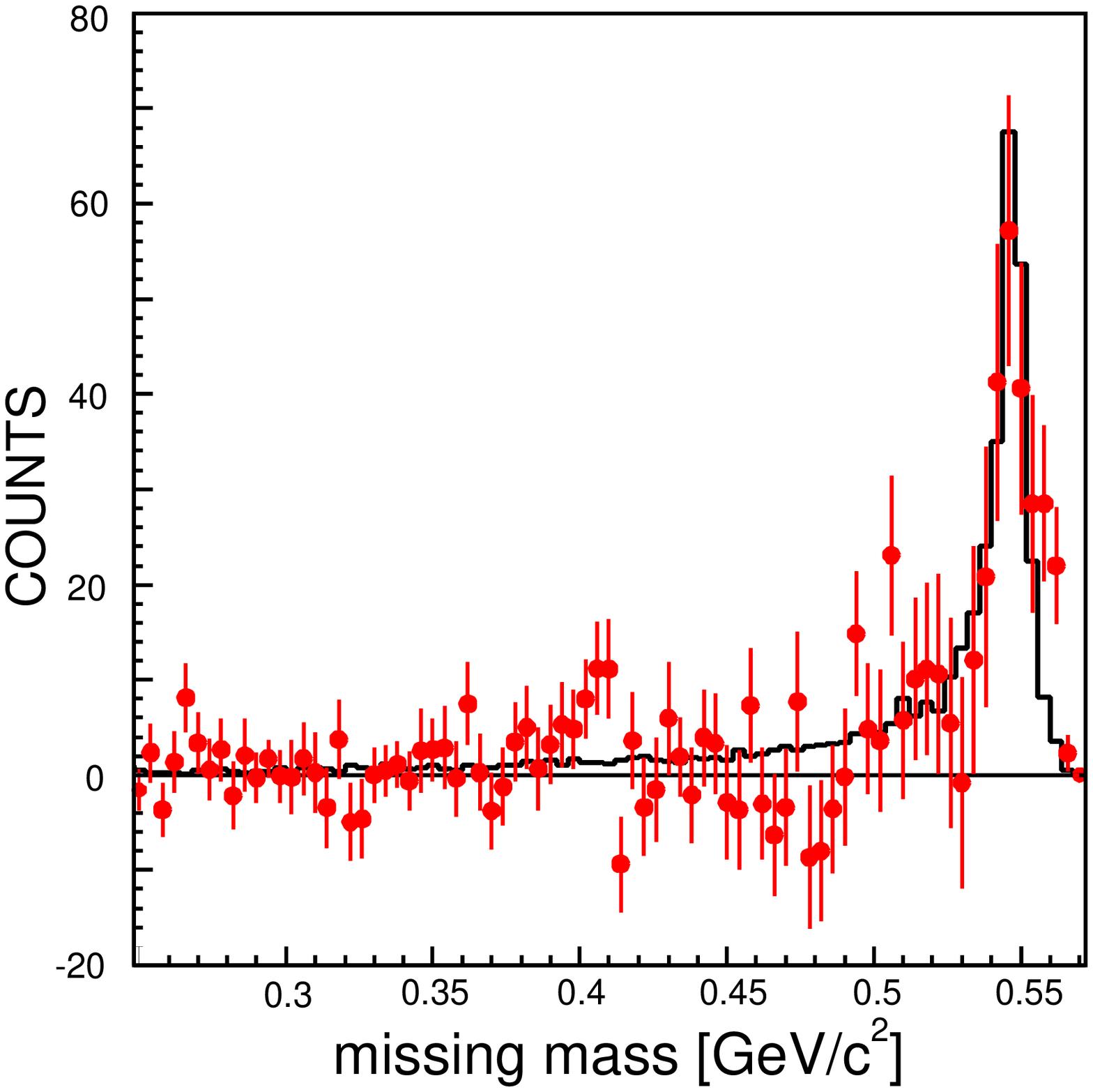,width=0.415\textwidth,angle=0}}}

       \vspace{-0.2cm}

       \parbox{0.55\textwidth}{\mbox{} }\hfill
       \parbox{0.41\textwidth}{a) }\hfill
       \parbox{0.02\textwidth}{b) }
\caption{\label{mmexp1}
          a) The solid line shows the missing mass distribution of the $pn\to pn X$
             process determined for $Q>0$ with respect to the
             $pn\to pn\eta$ reaction. 
             The sum for all $\Delta{Q}$ intervals is shown.
             The shaded histogram depicts the
             missing mass spectrum determined for $Q<0$. The procedure of
             the background construction is as described in the text.
          b) The points denote the experimental missing mass spectrum for $Q> 0$
             after subtraction of the multi-pion background.
             The superimposed solid line, normalised in amplitude to the data points,
             results from a Monte-Carlo simulation.
        }
\end{figure}

Recently the $\eta$ and $\eta^{\prime}$ mesons production in the
proton-neutron collisions has been studied close to the
kinematical threshold at the COSY-11
facility~\cite{hadron,joannaleap,michalproc}. Although the
evaluation of these data is still in progress, the first trials to
distinguish between the multi-pion background and the $\eta$ meson
production in the missing mass spectra lead to a clear signal for
the $pn\to pn\eta$ reaction. Fig.~\ref{mmexp1}a shows the sum of
the missing mass spectra for all $Q>0$ bins (solid line) and the
sum of the corresponding background spectra (shaded histogram).

The data were evaluated as follows. First the sample of events
with excess energy greater than zero was divided into a few
intervals of $\Delta{Q}~=~5~MeV$ and a missing mass spectrum was
calculated for each sub-sample separately. Next, for each of these
spectra, determined in interval $\Delta Q$ around a given
$Q_{\eta}$ value, a background was constructed from all events
with $Q$ less than zero. The missing mass of each background
event, measured at given excess energy $Q_{b}$, was increased by an
amount $(Q_{\eta} - Q_{b})$, as depicted in Fig.~\ref{mmpisum} and
described in section~\ref{estimation}. For each $\Delta{Q}$
interval separately, the background spectrum was normalised to the
distribution at a given $Q_{\eta}$ for mass values less than
0.25~GeV/c$^2$. 
In this region the events correspond to a single
pion production, and since the energy is high above the threshold
the cross section for $\pi^0$ production stays nearly constant
when the excess energy changes only by  few tens of MeV.

\begin{figure}[H]
       \vspace{-0.2cm}
       \parbox{1.0\textwidth}{\centerline{
                \epsfig{file=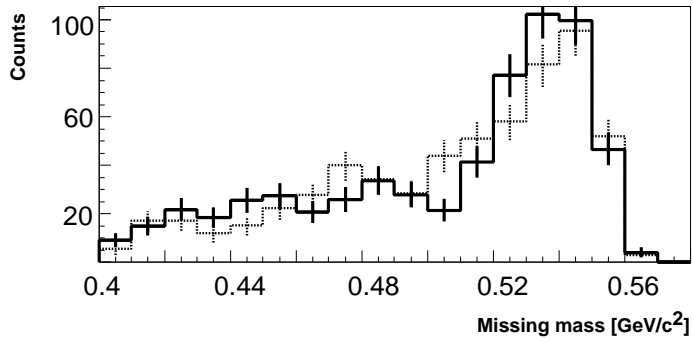,width=0.7\textwidth,angle=0}}}\\
       \vspace{-0.4cm}
   \parbox{1.0\textwidth}{\caption{ \small
           Background distribution obtained using events with Q ranging from
           -9~MeV to 0~MeV (dotted line) and from -40~MeV to -9~MeV (solid line).
           \label{monte7}
         }}
  \end{figure}

In order to make a check of the background extraction we have 
constructed it 
      dividing the range of Q values into two parts equal in number
      of counts. The comparison of the spectra resulted from the two parts is shown in 
      Fig.~\ref{monte7}. Indeed, the determined spectra from different ranges of Q
      are consistent within the statistical errors.

Fig.~\ref{mmexp1}b depicts the spectrum of the
missing mass for $Q>0$ after the background subtraction. The
superimposed line indicates the expectations from the Monte-Carlo
simulations, taking into account the momentum spread of the
beam~\cite{NIMbeamtarget}, Fermi momentum of the nucleons inside
the deuteron~\cite{lacombe1,lacombe2,rafalmgr}, beam and target
dimensions~\cite{NIMbeamtarget}, multiple scattering in the
detector components and air, and position-,
 time- and energy-
resolution of all detector
components~\cite{brauksiepe,NIMjurek,joannamgr,hab,magnusphd}.
Clearly, within the statistical uncertainties, the results of
simulations conform very well to the shape of the observed
distribution.

\begin{figure}[H]
        \parbox{0.14\textwidth}{\mbox{}}
        \parbox{0.43\textwidth}{\centerline{
                \epsfig{file=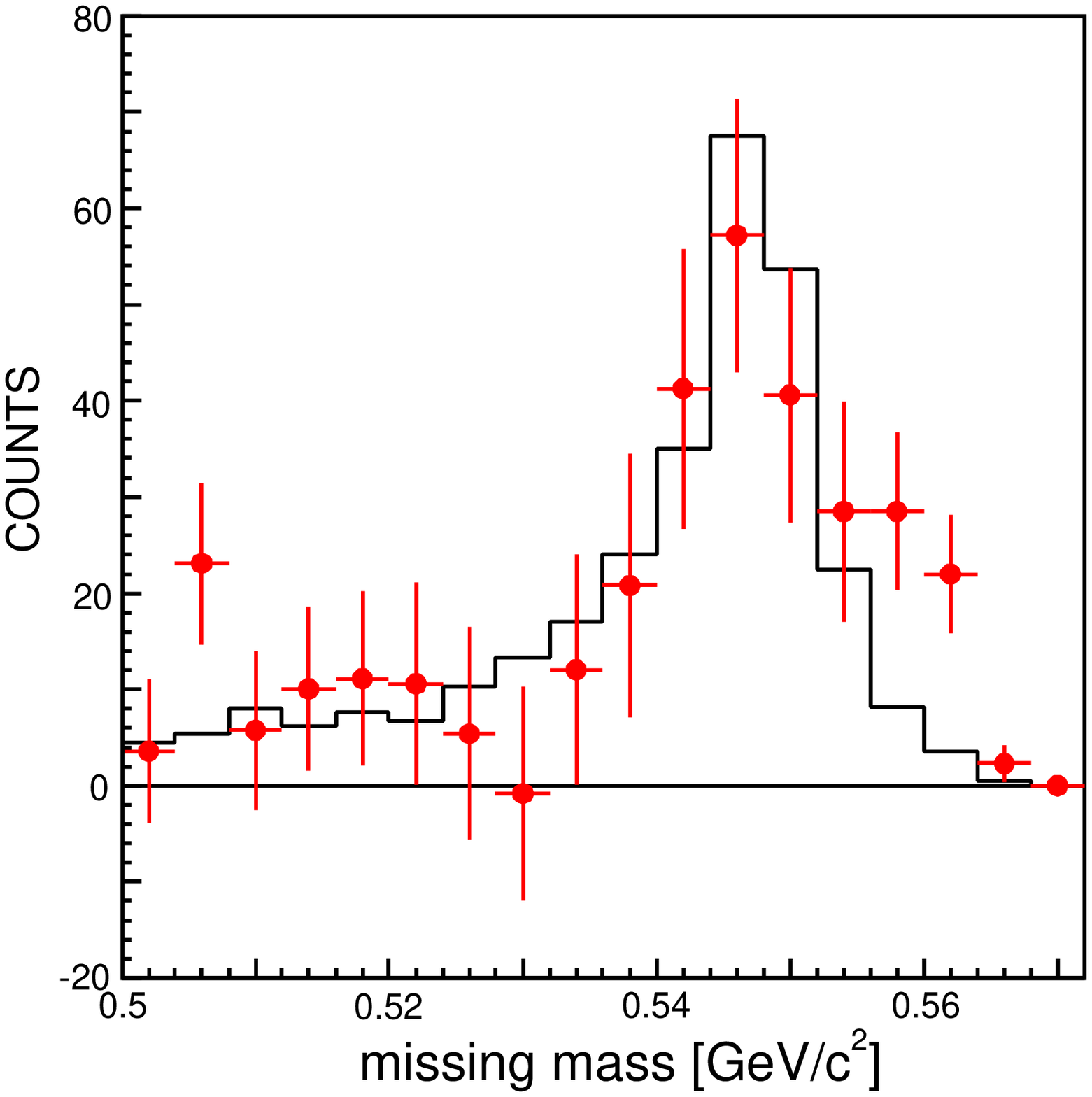,width=0.42\textwidth,angle=0}}}
       \parbox{0.43\textwidth}{\centerline{
                \epsfig{file=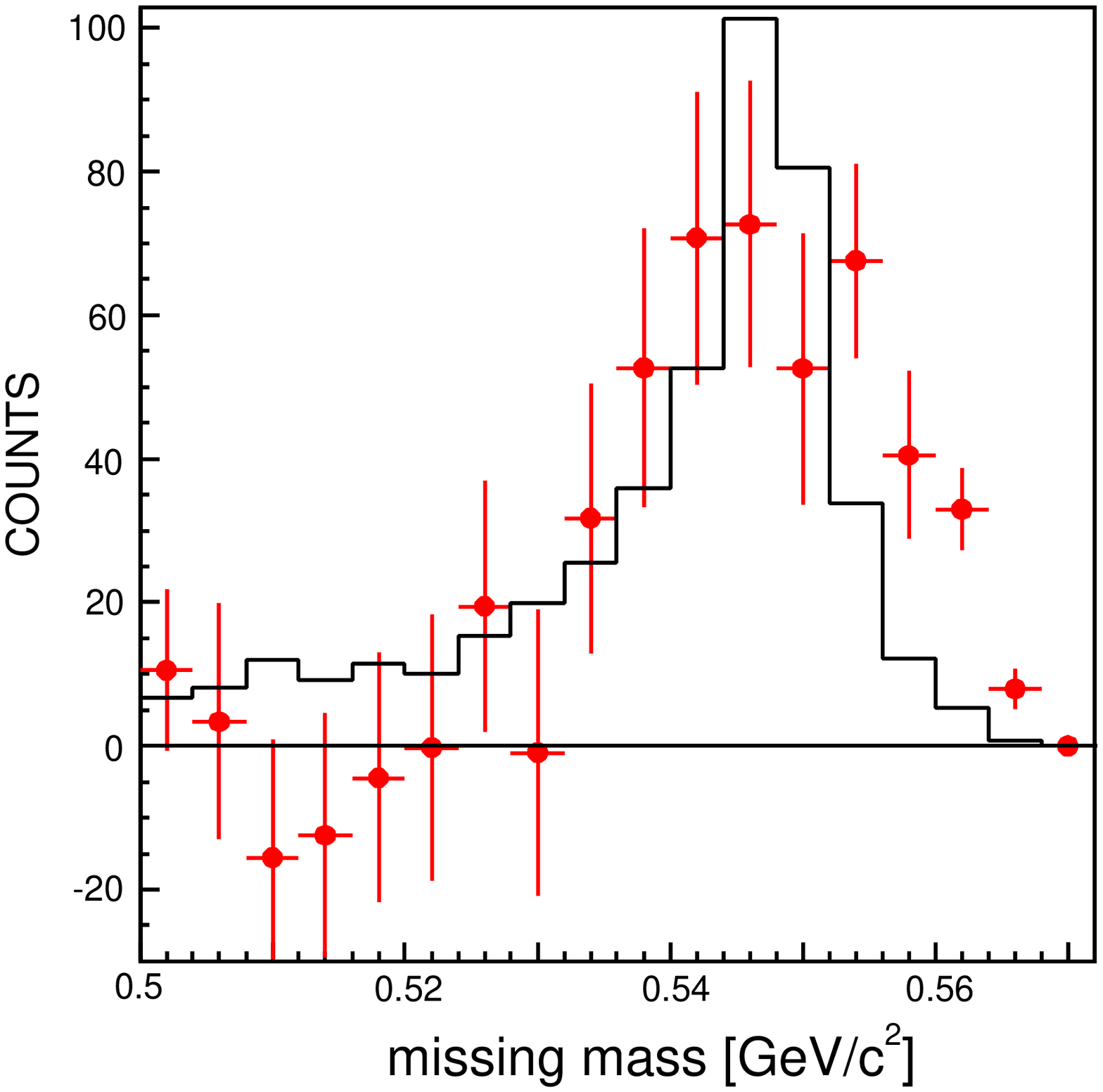,width=0.42\textwidth,angle=0}}}

       \vspace{-0.3cm}

       \parbox{0.55\textwidth}{\mbox{} }\hfill
       \parbox{0.43\textwidth}{a) }\hfill
       \parbox{0.02\textwidth}{b) }
       \vspace{-0.4cm}
\caption{\label{mmexp2}
         Missing mass spectra:  a) determined according to the method
         described in this article, and b) using an earlier approximate procedure~\cite{hadron}.
         The solid line depicts results of the simulation normalised to the data.
        }
\end{figure}

\begin{figure}[H]
       \vspace{-0.2cm}
       \parbox{1.0\textwidth}{\centerline{
                \epsfig{file=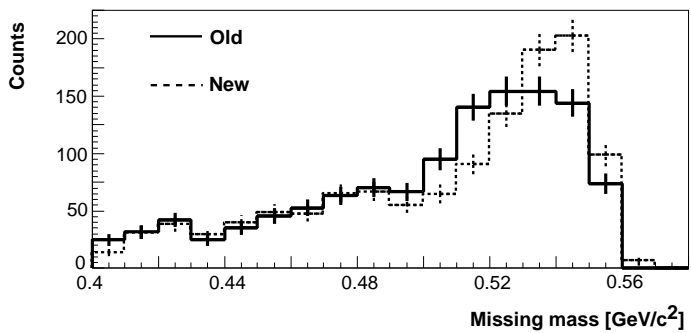,width=0.7\textwidth,angle=0}}}\\
       \vspace{-0.4cm}
   \parbox{1.0\textwidth}{\caption{ \small
         Form of the background determined according to the method
         described in this article (dotted line), 
         and using an earlier approximate procedure~\cite{hadron} (solid line).
           \label{monte9}
         }}
  \end{figure}

\begin{figure}[H]
       \vspace{-0.2cm}
       \parbox{1.0\textwidth}{\centerline{
                \epsfig{file=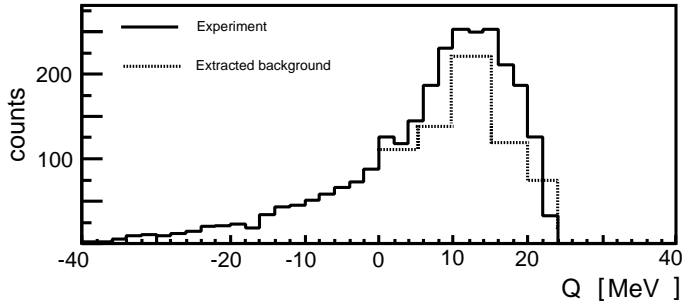,width=0.7\textwidth,angle=0}}}\\
       \vspace{-0.4cm}
   \parbox{1.0\textwidth}{\caption{ \small
           The solid line  shows experimental distribution of the excess energy Q 
           as measured with the COSY-11 detection setup (see also figure~\ref{mmpisum}),
           and the superimposed dotted histogram indicates a  spectrum of $dN^{\pi}/dQ(Q)$
           extracted for positive values of Q taking into account all events of $Q~<~0$ and 
           the method introduced in this article.
           \label{monte10}
         }}
  \end{figure}

For a better visualisation, the sums for all $\Delta{Q}$ intervals
are shown in Figs.~\ref{mmexp1}a and~\ref{mmexp1}b because the
individual spectra are statistically much less significant, due to
the small number of events in this first measurement, which was
meant only  as a feasibility test of the COSY-11 facility for the
study of the meson production via quasi-free proton-neutron
scattering. A more detailed view of the region around the $\eta$
peak is shown in Fig.~\ref{mmexp2}a. For comparison we present
also in Fig.~\ref{mmexp2}b the result obtained before the current
method was elaborated~\cite{hadron}. This first analysis was done
by deriving a missing mass background spectrum from all events
with $Q<0$ without adding (event-by-event) the difference of excess energies
between a corresponding background event and that from the bin for
which the signal of $\eta$ meson was being studied.
Although at first sight both spectra look similar, on closer
inspection one recognizes that the left hand histogram agrees
statistically much better with the experimentally extracted data
than that on the right. The difference is understandable since the
shift of the background to the kinematical limit diminishes the
number of background events on the left of the eta-peak and
increases that on the right (see Fig.~\ref{monte9}), 
leading to the better agreement with the
shape predicted by the Monte-Carlo studies. 
Moreover, the total
number of events assigned to the $pn\to pn\eta$ reaction, when
extracted according to the new procedure, is by about 30\% lower
than previously estimated~\cite{michalproc} and now conforms better to
the results obtained by the WASA/PROMICE
collaboration~\cite{calen1}, where photons from the decay of
the $\eta$ were detected.
Finally in Fig.~\ref{monte10}
  we present a $dN^{\pi}/dQ$ 
      function determined for the positive values of Q.
      One can see that the distribution $dN^{\pi}/dQ$ for positive values
      of Q is rather not constant within the $\Delta{Q}$ bin contrary to the
      assumption made when constructing the background.
      Therefore,  we have made an iterative
      procedure and construct a background taking in formula~\ref{equation2} the obtained
      distributuion of $dN^{\pi}/dQ$. The result was statistically the same even
      after the first iteration. This can be understood since the width of the
      $\Delta{Q}$  bin is rather small compared to the mass resolution. 
      One bin at the mass spectrum (eg. in figure~\ref{mmexp2} ) corresponds to 4 MeV.
      Therefore, even though the $dN^{\pi}/dQ$ significantly differs from being constant
      it hardly influences the $dN/dm$ spectrum  within the present statistics.
      A larger sample of data  will enable to determine $dN^{\pi}/dQ$ iteratively
      more precisely and hence will permit to obtain results not biased by the assumption that
      $dN^{\pi}/dQ$ is constant within a studied bin.

\section{Summary}
\label{summary}

A technique for disentangling contributions of multi-pion and
single meson production in the missing mass spectrum of quasi-free
$pn\to pnX$ reactions has been presented. The method enables one
to extract a signal for the $pn\to pn$~\emph{meson} reactions
measured close to the threshold, when the mass resolution is
comparable with the centre-of-mass excess energy. In this energy
range the peak in the missing mass spectrum from the single meson
creation is close to the kinematical limit and the separation of
the signal due to the single meson from that of the continuous
distribution associated with multi-pion production becomes
nontrivial. It was also shown how to combine events of different
excess energies in order to gain statistical significance in the
determination of the shape of the multi-pion mass distribution.

In the case of the $pd\to pnXp_{sp}$ reaction measured by the
COSY-11 collaboration~\cite{hadron}, we demonstrated that the
method leads to the derivation of the signal from the quasi-free
$pn\to pn\eta$ process that is consistent with expectations. This
successful application, and the fact that the technique is
independent of the detection system, suggests that it can be
applied more generally for distinguishing between single and
multi-meson production in missing mass spectra obtained in $pd\to
pnXp_{sp}$ reactions.

The presented method will be used for the extraction of the signal
from data on $\eta^{\prime}$ meson production via the $pn\to
pn\eta^{\prime}$ reaction~\cite{joannapnetap}, whose rate is
expected to be about forty times lower than that for the $\eta$
meson. These cross sections are utterly unknown and their
determination may be of crucial importance for the study of the
gluonic degrees of freedom in the $\eta$ and $\eta^{\prime}$ meson
production processes~\cite{steven2}.

{\bf Acknowledgments}

We appreciate very much  a helpful communication with Colin Wilkin
and we are grateful for his valuable correction of the manuscript.
We acknowledge the support of the
European Community-Research Infrastructure Activity
under the FP6 "Structuring the European Research Area" programme
(HadronPhysics, contract number RII3-CT-2004-506078),
of the FFE grants (41266606 and 41266654) from the Research Centre J{\"u}lich,
of the DAAD Exchange Programme (PPP-Polen),
of the Polish State Committee for Scientific Research
(grant No. PB1060/P03/2004/26),
and of the
RII3/CT/2004/506078 - Hadron Physics-Activity -N4:EtaMesonNet.

\end{document}